\documentclass[10pt,conference]{IEEEtran}
\IEEEoverridecommandlockouts
\usepackage{cite}
\usepackage{amsmath,amssymb,amsfonts}
\usepackage{algorithmic}
\usepackage{booktabs}
\usepackage{xcolor}
\usepackage{nicematrix}
\usepackage{enumitem}

\usepackage{url}

\usepackage{breakurl}
\usepackage[breaklinks]{hyperref}

\usepackage{multirow}
\usepackage{graphicx}
\usepackage{textcomp}
\usepackage{siunitx}
\usepackage[capitalize]{cleveref}
\def\BibTeX{{\rm B\kern-.05em{\sc i\kern-.025em b}\kern-.08em
    T\kern-.1667em\lower.7ex\hbox{E}\kern-.125emX}}
\long\def\longcaption#1#2{\centering\begin{minipage}{#1}\vspace{-0.7\baselineskip}\footnotesize\noindent\emph{#2}\end{minipage}}
\definecolor{Gray}{gray}{0.9}

\usepackage{array,booktabs,ragged2e}
\newcolumntype{R}[1]{>{\RaggedLeft\arraybackslash}p{#1}}

\usepackage{pifont}
\usepackage[T1]{fontenc}
\usepackage{pbsi}
\newcommand{\cmark}{\ding{51}}
\newcommand{\qmark}{\bsifamily ?}

\usepackage{tikz}
\usetikzlibrary{shapes}
\usetikzlibrary{patterns}
\usetikzlibrary{positioning}
\usepackage{pgfplots}
\pgfplotsset{compat=1.16}

\usepackage{caption}
\usepackage{subcaption}

\begin{document}


\title{A Benchmark Comparison of Python Malware Detection Approaches}


\author{\IEEEauthorblockN{Duc-Ly Vu}
\IEEEauthorblockA{\textit{Chainguard and FPT University} \\
lyvd@fe.edu.vn}
\and
\IEEEauthorblockN{Zachary Newman}
\IEEEauthorblockA{\textit{Chainguard} \\
zjn@chainguard.dev}
\and
\IEEEauthorblockN{John Speed Meyers}
\IEEEauthorblockA{\textit{Chainguard} \\
jsmeyers@chainguard.dev}
}

\maketitle

\begin{abstract}
While attackers often distribute malware to victims via open-source, community-driven package repositories, these repositories do not currently run automated malware detection systems.
In this work, we explore the security goals of the repository administrators and the requirements for deployments of such malware scanners via a case study of the Python ecosystem and PyPI repository, which includes interviews with administrators and maintainers.
Further, we evaluate existing malware detection techniques for deployment in this setting by creating a benchmark dataset and comparing several existing tools, including the malware checks implemented in PyPI, Bandit4Mal, and OSSGadget's OSS Detect Backdoor.

We find that repository administrators have exacting technical demands for such malware detection tools.
Specifically, they consider a false positive rate of even 0.01\% to be unacceptably high, given the large number of package releases that might trigger false alerts.
Measured tools have false positive rates between 15\% and 97\%; increasing thresholds for detection rules to reduce this rate renders the true positive rate useless.
In some cases, these checks emitted alerts more often for benign packages than malicious ones. 
However, we also find a successful socio-technical malware detection system: external security researchers also perform repository malware scans and report the results to repository administrators.
These parties face different incentives and constraints on their time and tooling.
We conclude with recommendations for improving detection capabilities and strengthening the collaboration between security researchers and software repository administrators.
\end{abstract}


\section{Introduction}
\noindent
After the creation of the first computer worm, the first malware detection software followed almost immediately~\cite{corewar}.
The prevalence of malware has only increased over time~\cite{sonatype-report-2021}.
Since then, entire textbooks~\cite{christodorescu2007malware, mohanta2020malware} and conferences~\cite{ieee-malware-proceedings} have been devoted to the automated detection of unwanted software, in the hopes of removing it from machines or preventing it from arriving in the first place.

Malware detection may have a role to play in overcoming software supply chain attacks. In such attacks, malicious software comes not from a compromise of a vulnerable system but rather from the development, building, and deployment of software~\cite{torres2019toto}.
These attacks can take many forms, from a formerly-benign software dependency incorporating malicious behavior after an update to ``trusting trust'' attacks~\cite{thompson1984reflections} in which a malicious compiler inserts a backdoor into the software it compiles (including itself). An effective malware detection system could halt the propagation of such attacks up the open-source software (OSS) supply chain.

Programming language-specific package repositories (such as npm, PyPI, or RubyGems) are both a target for supply chain attacks and a site from which to mount defenses.
Developers use such repositories to manage their software’s dependencies but, in doing so, incur supply chain risk: recent academic work~\cite{zahan2022weak, duan2021towards, vaidya2019security} quantifies and enumerates attacks using these repositories.
Other academic works~\cite{taylor2020defending, wyss2022fork} propose defenses for specific classes of attacks, and repositories~\cite{borins2022top} and nonprofit organization~\cite{ingram2022your} are deploying techniques like software signing and multifactor authentication.

However, deployments of malware-detection solutions in these settings have met with limited success in integrating into a package repository. To the best of our knowledge, none of the current techniques have been successfully adopted by PyPI. These repositories differ from the traditional settings for such malware scanners: antivirus software and intrusion detection systems. Consequently, techniques designed for those settings may not be suitable for a package repository.
For instance, in 2020, the Python Software Foundation oversaw the creation of a malware scanning pilot (called PyPI malware checks) for Warehouse, the application which serves PyPI~\cite{pypi-malware-checks}. However, two years later, administrators disabled its malware checks.

\subsection{Objectives and methods}
\noindent
In this work, we seek to understand the requirements for deploying malware detection tools in the package repository setting as well as the fitness of current malware detection techniques for this application.
To do so, we study the Python ecosystem and the PyPI repository, focusing on Python-specific malware detection techniques at scale.
We conduct interviews (see \cref{sec:interviews}) with contributors to and maintainers of PyPI, along with a software repository security researcher.
In \cref{sec:evaluation}, we create a Python malware benchmark dataset and evaluate existing automated techniques. We analyze the results of the experiments in Section \ref{sec:discussion}.

\subsection{Findings and recommendations}
\noindent
In our user research (\cref{sec:interviews}), we find that the motivation and even the setting for repository-side malware scanning differ in practice from those assumed in the literature.
Maintainers want to mitigate low-effort, high-volume attacks with minimal effort of their own, not remove all malware.
Additionally, repository administrators possess demanding criteria for a practical deployment of repository-run checks; false positive rates must be virtually zero.
The benchmarking analysis reveals that all the current automated detection methods unambiguously fail to meet these criteria, exhibiting false positive rates of \SI{15}{\percent} or higher.
Instead, the interviews revealed that the true ``malware checks'' in the PyPI ecosystem involve external researchers with their own incentives, who find and report malware to the maintainers~\cite{pypi-security-reporting}.

We propose recommendations for researchers and repository maintainers (\cref{sec:recommendations}) in order to improve the security of these ecosystems, especially for malware detection. In short, most researchers should primarily focus on other efforts to secure package repositories in collaboration with the maintainers, who are in the best position to understand their security needs and have more ideas than time.
In automated malware detection, researchers should focus on aligning the incentives of repository maintainers and external researchers, and engage with malware detection as a human-computer system.
All parties should attempt to facilitate smoother collaboration and deployment with better tooling and data.

\subsection{Contributions} This work contributes:
\begin{itemize}[leftmargin=0.3cm]
    \item Qualitative analysis of the security goals and logistical requirements of the main Python package repository.
    \item A benchmark data set for comparing malicious and benign Python packages.
    \item A cross-comparison of recent automated Python malware-detection tools.
    \item Recommendations for both researchers and repository maintainers.
\end{itemize}

\subsection{Security and ethical considerations}
\noindent
This work adheres to the ACM Publications Policy on Research Involving Human Participants and Subjects~\cite{acm2021policy}.
All interviewees gave informed consent prior to their interviews and approved the characterization of their responses presented herein; they received a copy of our findings and the opportunity to present feedback before submission. They are identified only by their background and not their name or institutional affiliation.
Even still, it may be possible to identify them from unique details; the interviewees acknowledge this risk and consider themselves ``public figures.''

The malware samples used for analysis are maintained by their collectors~\cite{backstabbers-homepage, maloss-homepage}, and are not publicly available (but are available on request by the respective researchers) to prevent their misuse.

\section{Background}
\noindent
Software supply chain attacks have been on the rise over the past ten years~\cite{trey2020breaking, geer2020good}.
In particular, attackers have increasingly exploited the trust that software developers and downstream consumers have implicitly placed on open-source infrastructure, especially package repositories.
Fortunately, repository administrators and a number of allied parties have begun implementing defenses against attacks on the open-source software supply chain, including the automated detection of malicious packages.
However, despite decades of research and engineering on general automated malware detection software, open-source software malware detection in this setting is at an early stage in both theory and practice.
The nascent state is due in part to a lack of consistent benchmarking and to differences between the goals of malware detection researchers and the needs of repository maintainers.

\subsection{Software supply chain attacks}
\noindent
The interval between Ken Thompson’s theoretical discussion of backdoored compilers~\cite{thompson1984reflections} to arguably the first actual software supply chain attacks was nearly twenty years.
These days, software producers and consumers are fortunate when there are twenty days between known attacks, not to mention the countless unreported attacks~\cite{geer2020good, iqtlabs-dataset}.
In recent years, notable attacks have included, for instance, SolarWinds, in which the Russian intelligence services injected malware into a widely used network management software~\cite{gao2022federal}.
Malicious attacks on the software supply chain have also targeted open-source software producers and consumers.
One particularly notable attack was on consumers of event-stream, a popular JavaScript package.
In 2018, a new open-source software maintainer took control of the package and introduced malicious code to exfiltrate information from Bitcoin wallets.
This package averaged nearly two million downloads per week and went undiscovered for nearly two months~\cite{arvanitis2022systematic}. 

\paragraph{Open source repositories have been targeted}
Open-source software repositories, essentially stores of free and open-source (FOSS) components that facilitate code reuse, have become an important part of the software supply chain over the past couple of decades.
Developers use such repositories to search, add, and manage dependencies for the software they write, which allows them to write more complicated software more quickly.
For instance, the Python Package Index or PyPI~\cite{pypi-homepage}, as of August 2022, contains almost \num{400000} projects and nearly \num{4000000} releases. PyPI, like many registries, is governed by a non-profit foundation and is run by a handful of volunteers~\cite{warehouse-application}.

Attacks on the software supply chain have, unfortunately, but perhaps inevitably, therefore extended to open-source repositories and their users.
Critics argue that such repositories facilitate overreliance on external dependencies, which in turn increases the attack surface for a supply chain attack.
This ``overreliance'' from the critic’s perspective also increases the leverage of an attacker able to compromise a popular package~\cite{haney2017npm}. 

There are a number of classes of attacks on open-source package repositories.
Typosquatting, in which an attacker relies on a user mistyping the name of a more popular package or confusing one package for another, has begun to receive analytical attention~\cite{vu2020typosquatting, taylor2020defending}.
So have account takeover attacks in which an attacker steals credentials or otherwise gains control over a package and then alters the package code~\cite{sharma2022pypi, ohm2020backstabber}.
Relatively less appreciated are the threats of ``shrinkwrapped clones''~\cite{wyss2022fork}, which duplicate existing packages in order to attract downloads before introducing malware, and ``domain resurrection'' attacks in which an attacker uses an abandoned email domain to take control of a package~\cite{zahan2022weak}.
Attackers have compromised the repositories themselves, and not just individual packages~\cite{cappos2008look, iqtlabs-dataset}.
Recent work has created a comprehensive taxonomy of attacks on the open-source software supply chain~\cite{ladisa2022taxonomy}. 

Malicious packages can deploy malicious code either at installation time or runtime.
Repository-based malware often exploits install-time behavior, injecting malicious code into install scripts associated with a package~\cite{ohm2020backstabber}.
This is often the case with Python malware, as Python packages can execute arbitrary code on installation via \texttt{setup.py} files.
It is also possible for malicious Python packages to exhibit runtime malicious behavior (e.g., by importing a malicious dependency)~\cite{ohm2020backstabber, lutoma2019dateutil}.
Recent cases of ``protestware'' also fit into this category~\cite{newman2022fragile, kula2022war}. 

\paragraph{Repository defenses against supply chain attacks}
Repository maintainers have not stood idly by while attacks on registries have increased.
Recently some major repositories have implemented two-factor authentication for the maintainers associated with widely used packages as a way to reduce the likelihood of consequential account takeovers~\cite{sharma2022pypi, anderson2022rubygems}.
Software signing is a promising technique for preventing attackers from deploying software via compromised accounts.
Some repositories are considering integration with Sigstore~\cite{sigstore}, a project that enables trusted parties to make authenticated claims about software artifacts~\cite{hutchings2022new}.

These recent developments complement earlier efforts, some still ongoing, to secure repositories (including PyPI) using technologies like The Update Framework (TUF) to enable recovery in the event of compromise~\cite{samuel2010survivable, pep458, pep480}. Reproducible builds~\cite{reproducablebuilds}, a set of techniques to verify
that no vulnerabilities or backdoors have been introduced
during the build process, is promising but appears hard to achieve for interpreted languages such as Python~\cite{vu2021lastpymile} or JavaScript~\cite{goswami2020investigating}.

Repository maintainers and the communities associated with key repositories have also experimented with active scanning for malware~\cite{package-analysis-tool, ossgadget}. The next subsections discuss the current status of malware detection, focused on OSS repositories. 

\subsection{Malware detection in PyPI and other Repositories}
\noindent
All supply chain attacks have a goal of distributing unwanted software to unwitting downstream dependents, so an automated system that could detect unwanted software could prevent such attacks.
The automated malware detection literature, dating back decades, attempts to do exactly that; we examine the approaches most relevant to the package repository setting.

\paragraph{Automated malware detection}
Malware detection technologies aim to identify malicious or unwanted software in an automated way~\cite{idika2007survey, aslan2020comprehensive, ye2017survey}.
Typical use cases include antivirus software or intrusion detection systems.
Some methods are signature-based: they try to match specific malware (for instance, by checking file hashes against a blocklist).
Although signature-based approaches (different from the signatures in software signing) can catch specific malware precisely, they do not generalize to new malware samples, even when they are very similar to old ones.

Other methods are behavior- or anomaly-based, and attempt to detect suspicious patterns or behaviors in software~\cite{bazrafshan2013survey}.
Behavior-based methods include static and dynamic analysis.
Static analysis techniques analyze software without running it---for instance, looking for suspicious imports in the PE header of a binary.
Dynamic analysis techniques execute the software (typically in an isolated sandbox environment, for security) to observe its actual behaviors at runtime.
These approaches generalize to unseen malware but risk false positives.

\paragraph{Software repository malware detection}
In the literature, many approaches have been proposed to identify malicious packages in package repositories such as npm or PyPI~\cite{duan2021towards, sejfia2022practical, ohm2022feasibility, vu2020towards} or in the source code repositories such as Github~\cite{gonzalez2021anomalicious, cao2022fork}.
Most such approaches analyze different aspects of a package using metadata~\cite{vu2020typosquatting, taylor2020defending}, static~\cite{pypi-malware-checks, ossgadget-oss-detect-backdoor, bandit4mal, bertu2018detecting}, or dynamic~\cite{duan2021towards, package-analysis-tool, wright2020hunting} analysis.
These approaches use indicators such as sensitive API calls or abnormal network connections borrowed from traditional malware detection techniques~\cite{sikorski2012practical}.
Package repositories usually keep the metadata of a package (e.g., package name, author names).  Metadata analysis techniques examine those metadata to flag suspicious packages. For example, using package names and package popularity to flag typosquatting or combosquatting packages~\cite{vu2020typosquatting, taylor2020defending}.
However, these research projects often present their work in the context of potential deployment by open-source repository maintainers.

Commercial OSS malware scanners such as Sonatype's automated malware detection system~\cite{sonatype-automated-detection} continuously monitor newly added packages for their maliciousness.
However, commercial security companies often do not open-source or otherwise release their detection scheme, which prevents researchers from evaluating these approaches.

\paragraph{A gap between theory and practice}
Many researchers advertise their malware detection techniques as a way to keep unwanted packages out of a repository on an ongoing, automated basis, with high true-positive rates as evidence of their tool’s efficacy.
However, such tools rarely acknowledge the unique requirements of the setting.
The most important criterion is that a check must provide a ``useful signal,'' and a check that produces many false positives should be removed~\cite{pypi-malware-checks}.
We observed that there is no research to understand if the requirements of a useful malware check by PyPI administrators would be satisfied by either academic or industrial tools. This motivates the next exploratory study. 

\section{What Do Experts Say About Scanning PyPI for Malware?} \label{sec:interviews}
\noindent
To better understand the challenges in scanning malware for open-source software repositories, and PyPI in particular, this study included a qualitative, interview-based, user research component.
Individuals in the PyPI maintainer community, security researchers, and academics already have firsthand experience in scanning open-source packages for malware, and so this study sought to benefit from their first-hand knowledge and better appreciate the promise and challenges of malware scanning in the context of an open-source software repository~\cite{taylor2020defending, wyss2022fork}. 

\subsection{User research interview methodology}
\noindent
The research team conducted three interviews in July and August 2022. This section describes the objectives, the questionnaires and interview format, and the main findings.

\paragraph{Research Objectives}
We initially sought to answer a narrow question: what are the performance properties of a technical system for malware detection in the context of a large OSS repository?
This information was intended to provide input on what technical criteria ought to be used to distinguish practical and impractical approaches to malware detection for PyPI and other similar registries.
The goal was to directly aid the PyPI maintainers in assessing and improving a malware check system to be operated by the open-source software maintainers associated with PyPI.
The research objective evolved as the interviews revealed new perspectives on OSS malware detection.

While interviewees did provide answers to the earlier, more narrow question, the research team ultimately settled on a broader question: what should be the characteristics of a socio-technical system that enables malware detection in the context of a large OSS repository?
This framing shifted responsibility for malware detection solely from the shoulders of PyPI maintainers to a combination of PyPI maintainers, security researchers, and academics.
Importantly, this perspective also de-emphasizes the technical characteristics of any particular scanning approach and highlights the need for multiple parties to perform malware scanning.
More on this perspective will be included in the key findings section (\cref{sec:interview-key-findings}).

\paragraph{Interview Methodology} The interview questionnaire included five broad questions.
\begin{enumerate}[leftmargin=0.5cm]
    \item  What is the origin story of the current PyPI malware checks?
    \item What has been your experience, if any, with the current PyPI malware checks?
    \item What are the current plans, if any, for improving the PyPI malware checks?
    \item How do you judge the performance of a PyPI malware check system?
    \item How would you judge a set of proposed improvements to the PyPI malware check system?
\end{enumerate}

\noindent
In short, the interview questionnaire largely focused on answering the narrow questions consistent with the project’s initial research goal.
The interviews were, however, semi-structured, which allowed for the interviewers and interviewees to shift topics as new insights emerged.

All interviews were conducted over video teleconference and lasted approximately one hour.
The research team did not record the interviews but did take detailed notes.
The interviews were analyzed for common themes, which are presented below in the following section.

Because we chose to focus on a single ecosystem (PyPI), the number of individuals with direct experience scanning for malware was limited, which constrained our interviewee sample size. In particular, there are under ten people involved directly in the malware scanning activities in PyPI~\cite{warehouse-application}.

Two interviewees were drawn from the PyPI maintainer community and had direct experience with the PyPI malware system, either designing and building the system or operating it.
Another interviewee, an academic security researcher, had extensive experience with OSS malware detection.

\subsection{Interview key findings}\label{sec:interview-key-findings}
We distilled four main findings, one related to the narrow technical questions about current malware detection performance for PyPI malware checks and three related to broad questions about socio-technical system design related to OSS malware detection.

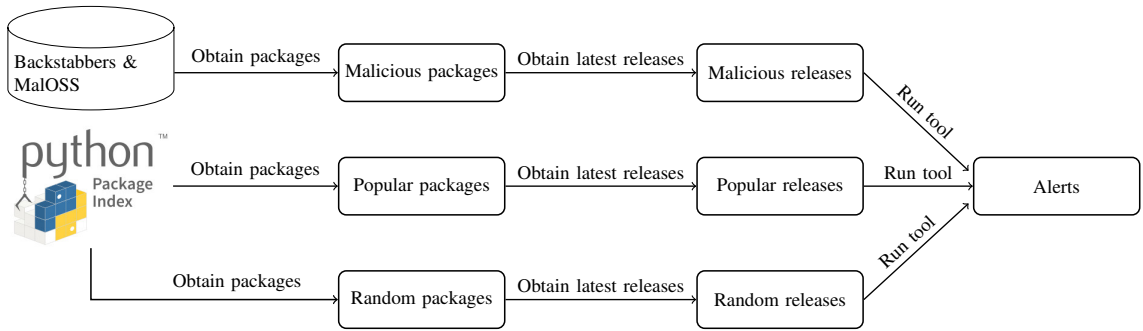
\begin{figure*}[t]
\centering
\resizebox{6in}{!}{
\begin{tikzpicture}
    \node (malicious-db) at (0, 0)  [cylinder, 
        shape border rotate=90, 
        draw,
        text width=8em,
        minimum height=1cm,
        minimum width=1cm,
        shape aspect=.25,] { Backstabbers \& MalOSS};
    \node (malicious-packages) [right=3cm of malicious-db, text width=8em, align=center, draw,thick, rounded corners,minimum height=3em, minimum width=6em]{Malicious packages};
    \node (malicious-releases) [right=3.5cm of malicious-packages, text width=8em, align=center, draw,thick, rounded corners,minimum height=3em, minimum width=6em]{Malicious releases};
    \node[inner sep=0pt, below = 0.25cm of malicious-db] (pypi-db) {\includegraphics[width=3cm, height=3cm, keepaspectratio,]{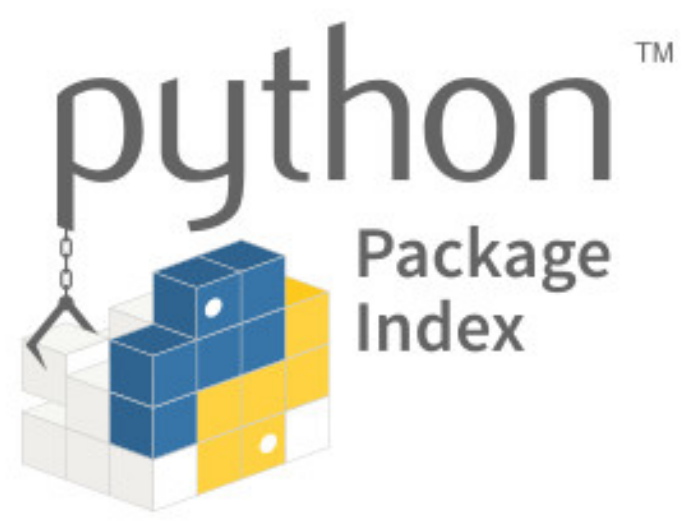}};
    \node (popular-packages) [below=1cm of malicious-packages, text width=8em, align=center, draw,thick, rounded corners,minimum height=3em, minimum width=6em]{Popular packages};
    \node (popular-releases) [below=1cm of malicious-releases, text width=8em, align=center, draw,thick, rounded corners,minimum height=3em, minimum width=6em]{Popular releases};
    \node (random-packages) [below=1cm of popular-packages, text width=8em, align=center, draw,thick, rounded corners,minimum height=3em, minimum width=6em]{Random packages};
    \node (random-releases) [below=1cm of popular-releases, text width=8em, align=center, draw,thick, rounded corners,minimum height=3em, minimum width=6em]{Random releases};
    \node (alerts) [right=2cm of popular-releases, text width=8em, align=center, draw,thick, rounded corners,minimum height=3em, minimum width=6em]{Alerts};

    \draw[->,thick, line width=0.25mm] (malicious-db) -- (malicious-packages) node[midway,above, sloped] {Obtain packages};
    \draw[->,thick, line width=0.25mm] (malicious-packages) -- (malicious-releases) node[midway,above, sloped] {Obtain latest releases};
    \draw[->,thick, line width=0.25mm] (malicious-releases.east) -- (16.1, -1.8) node[midway,above, sloped] {Run tool};
    \draw[->,thick, line width=0.25mm] (pypi-db) -- (popular-packages) node[midway,above, sloped] {Obtain packages};
    \draw[->,thick, line width=0.25mm] (popular-packages) -- (popular-releases) node[midway,above, sloped] {Obtain latest releases};
    \draw[->,thick] (pypi-db.south) |- (random-packages.west) node[midway,above, sloped]{};
     \node[text width=5cm] at(4, -3.85) {Obtain packages};
    \draw[->,thick, line width=0.25mm] (random-packages) -- (random-releases) node[midway,above, sloped] {Obtain latest releases};
    \draw[->,thick, line width=0.25mm] (random-releases.east) -- (16.1, -2.4) node[midway,above, sloped] {Run tool};
    \draw[->,thick, line width=0.25mm] (popular-releases.east) -- (alerts.west) node[midway,above, sloped] {Run tool};
\end{tikzpicture}
}
\caption{Diagram of Benchmarking Python Malware Detection Tools on the PyPI packages}
\label{fig:benchmarking-method}
\end{figure*}

\textbf{Current systems have too many false positives.}
The current PyPI malware checks produce an overwhelming number of false positives.
According to the interviewees associated with the PyPI maintainer community, this high rate of false positives reduces the usefulness of the system to such an extent that these checks have little to no value.
The academic interviewee also emphasized that his own OSS malware research had often encountered issues with high false positive rates and so cautioned that any PyPI malware scanning would similarly deal with this problem.

According to the interviewees, this high false positive rate does not, however, indicate that all OSS repository malware scanning is a wasted effort.
Both of the PyPI maintainers interviewed explicitly mentioned that PyPI’s experimentation with malware detection has been valuable, ultimately indicating the difficulty of the task.

\textbf{Repository administrators are overextended.}
The burden on a single party, especially PyPI maintainers directly, to scan PyPI for malware is prohibitively high~\cite{vu2020typosquatting, warehouse-application}.
Concretely, the interviewees report that an administrator is willing to spend about \num{20} minutes per week actively responding to reports of malware.
Consider that there are \num{20000} new PyPI releases each week~\cite{pypi-dataset-bigquery} and that administrators, according to our interviews, take down about \num{20} malicious packages in that same time period.
If an alert for a package takes about \num{1} minute to triage, even just the malicious packages consume the entirety of the time allotted for review of suspicious packages.
Given this, the interviewees were either reluctant or ambivalent about PyPI maintainers themselves processing the numerous alerts from an automated system in their relatively scarce volunteer time.

\textbf{Administrators have other security priorities.}
While, all things equal, these maintainers want to eliminate malware from PyPI, they nonetheless emphasized that only a small portion of PyPI malware actually harms users, noting that many malicious packages have almost no downloads after accounting for automated mirrors of the entire repository.
Further, the maintainers have little interest in the arms race of automated malware detection~\cite{waldo2018ending}: they believe that a sophisticated actor can always subvert such checks.
Instead, they prefer to use malware detection techniques only to eliminate frequent, low-effort attacks.
This allows them to focus on what they consider to be higher-impact security initiatives.

\textbf{A more practical approach to PyPI malware emphasizes collective protection.}
The currently-practiced approach to PyPI malware detection is a symbiotic relationship between PyPI and companies and individuals performing security research.
A set of researchers, some at software security companies and some as individuals, independently scan PyPI for malware and report malicious packages.
These parties, especially the software security companies, likely perform this scanning to gain attention for their companies, products, and services.
PyPI maintainers benefit because these parties maintain their own scanning infrastructure and undertake the tedious work of sifting through false positives themselves.

\subsection{Technical requirements for malware detection systems}
We asked interviewees about the requirements for an automated, repository-side malware scanning system.
First, they reported that the scanner must have a binary outcome, possibly via a threshold over a numerical score.
Approaches that simply return raw alert data and require ad-hoc data analysis, according to the interviews, are too time-consuming.
However, maintainers do prefer to manually review any packages they take down, so further details must be accessible.
Third, engineering resources are limited, though computational ones are less so.
Consequently, a scanner that executes code via dynamic analysis is unrealistic due to security concerns around maintaining a secure sandbox even though resources to run such a scanner at scale are available.
Second, any approach must effectively have a false positive rate of zero, though this is an extremely demanding requirement.

\section{How do malware detection approaches perform?} \label{sec:evaluation}

\noindent
In this section, we experimentally assess the reported false positives issue associated with Python malware detection that the interviewees discussed.
To do this, the research team created a benchmark dataset containing both malicious and benign PyPI packages and then evaluated the current PyPI malware checks in addition to two other selected Python malware detection approaches in \Cref{tab:tools}.

\subsection{Approach}
\noindent
\Cref{fig:benchmarking-method} depicts the benchmarking methodology.
We first chose a set of Python malware detection approaches to benchmark using the formal criteria discussed below.
We then collected both malicious and benign PyPI packages.
The benign dataset can be further subdivided into popular packages and random PyPI packages.
Finally, we scanned each package with each tool, recording all alerts produced.
An alert associated with a malicious package is considered a true positive, while an alert on a benign package is a false positive. 

\paragraph{Tool selection}\label{par:tools-selections}
We used the work of Ladisa et al.~\cite{ladisa2022taxonomy} and surveyed existing literature to create a list of candidate Python malware detection tools.
We used the following criteria for selecting tools to benchmark:
\begin{enumerate}[leftmargin=0.5cm]
    \item \textit{Behavior-based malware detection tools.} While several other tools (e.g.,~\cite{taylor2020defending, vu2020typosquatting}) analyze package metadata such as package name or package downloads, we focus on tools that analyze code.
    \item \textit{Source code available.} Our benchmarking analysis requires access to the tool’s source code. Specifically, the details about the detection technique (e.g., which kind of artifact or a file that a tool processes or ignores) must be available.
    \item \textit{Detection rules available.} Some tools, such as OSSF package-analysis~\cite{package-analysis-tool}, only provide  raw analytical results. Evaluating those tools requires a researcher to write their own detection rules, a potential source of analytical bias. Therefore, this benchmarking exercise only considered the tools that include detection rules. 
\end{enumerate}

\noindent
\Cref{tab:tools} contains the Python malware detection tools considered for inclusion in this study.
The tools range from simple heuristics and static analysis tools to complex dynamic analysis tools.
Most, though not all, are available on Github.
Three tools met all inclusion rules: the PyPI malware checks~\cite{pypi-malware-checks}, Bandit4Mal~\cite{bandit4mal}, OSSGadget's OSS Detect Backdoor~\cite{ossgadget-oss-detect-backdoor}.

\begin{table*}[t]
  \caption{Python malware detection tools considered for benchmarking.}
  \longcaption{\columnwidth}{The selected tool must satisfy all the conditions in the three columns. The question mark means the information of a criteria is unknown}
  \label{tab:tools}
  \begin{NiceTabular}{@{}lccc@{}}
    \toprule
    Approach & Source code available & Behavior-based detection & Detection rules available \\ \midrule
    Bandit4Mal~\cite{bandit4mal}  & \cmark & \cmark &   \cmark   \\ 
    OSSGadget's OSS Detect Backdoor~\cite{ossgadget-oss-detect-backdoor}  & \cmark  & \cmark & \cmark   \\
    PyPI Malware Checks~\cite{pypi-malware-checks}  & \cmark  & \cmark & \cmark \\ 
    Aura~\cite{carnogursky2019thesis}  & \cmark & \cmark & \\ 
    JFROG XRAY~\cite{jfrog-xray}  &  & \qmark & \\ 
    OSSF package-analysis~\cite{package-analysis-tool}  & \cmark  & \cmark & \\ 
    MalOSS~\cite{duan2021towards}  & \cmark  & \cmark & \\ 
    pypi-scan~\cite{pypi-scan}  & \cmark  & \cmark & \\ 
    Sonatype automated malware detection~\cite{sonatype-automated-detection}  &   & \qmark & \\
    TypoGard~\cite{taylor2020defending}  &   &  & \cmark \\ \bottomrule
  \end{NiceTabular}
\end{table*}

The PyPI malware checks were pushed in February 2020 by PyPA, as part of a modular pipeline capable of running many different types of malware analysis~\cite{warehouse-pr-7377}. 
The only deployed malware check, a ``setup pattern check,'' uses regular-expression-based YARA rules~\cite{yara} to scan \texttt{setup.py} files (which run on package installation).
This check includes four rules that alert suspicious imports and API calls in \texttt{setup.py} files. In our experiments, we used PyPI malware check rules in combination with a tool called yara-scanner~\cite{yara-scanner} to scan the packages.
As of 2022, repository administrators do not actively use the alerts generated by the checks in this pipeline.

OSS Detect Backdoor~\cite{ossgadget-oss-detect-backdoor} is a tool in a collection of tools called OSSGadget developed by Microsoft in June 2020. The collection provides \num{12} utilities to analyze different aspects of open source projects (e.g., health metrics).
OSS Detect Backdoor includes \num{41} rules checking that check for regular-expression-based malicious patterns (e.g., a reverse shell) of every text file in a package.
The tool currently supports scanning live packages from \num{15} package repositories, local packages, and generic URLs linked to packages.  

Bandit4Mal~\cite{bandit4mal} is a custom version of Bandit~\cite{bandit}. Bandit4Mal includes 45 rules to capture malicious patterns in a package specifically. 
Because Bandit4Mal relies on AST analysis, it must target a specific Python version. Our benchmarks use the Python 3 version of Bandit4Mal because of the wide use of Python 3 among Python developers~\cite{jetbrains2021survey}. 

The PyPI malware checks run only on \texttt{setup.py} files, on the theory that Python malware commonly runs at install-time and those suspicious behaviors are easier to catch in these files.
However, some creators of Python malware inject malicious code into other files executed at runtime~\cite{vu2021lastpymile}.
Therefore, in our evaluation (\Cref{sec:evaluation}), we first ran each tool only on \texttt{setup.py} files and then, second, on all Python files and recorded our evaluations separately.

\subsection{Benchmark dataset} \label{sec:experiments}
\noindent
Our benchmark dataset comprises malicious and benign Python packages collected from real-world attacks and PyPI.

\paragraph{A malicious Python package dataset}
Assembling a malicious Python package dataset was straightforward due to recent data collection efforts.
The Backstabber’s Knife Collection dataset~\cite{ohm2020backstabber, backstabbers-homepage} (commit \texttt{22bd76}) contains \num{107} Python packages previously identified as malware.
The MalOSS dataset~\cite{duan2021towards, maloss-homepage} (commit \texttt{2349402e}) contains \num{140} valid examples.
Both datasets have been collected manually from and represent real-world attacks.
The MalOSS dataset also includes malicious packages detected by the authors while scanning the PyPI repository. 

If a package was present in both datasets, the benchmark dataset includes only one instance.
The dataset also removed one duplicate of a similarly named package with the same contents.
In addition, the malicious dataset omitted three packages that do not contain any Python files.
Only the latest version of each package was included.
In total, the dataset contains \num{168} malicious Python packages.

\paragraph{Benign Python package datasets}
Ideally, this benchmark dataset would include a set of Python packages that is universally viewed as malware-free.
Unfortunately, there is no such existing dataset.
This paper, therefore, proposes two different second-best approaches.

\textit{Popular PyPI packages:}
We created a combined dataset of the \num{1000} Python packages~\cite{hugovk_top_pypi_packages} that are most downloaded and \num{1000} most widely depended upon Python packages.
Malware published to popular Python packages is typically removed very quickly, so the likelihood that these packages were malicious is low.
We excluded three popular packages that lack any Python files.
After deduplicating packages found in both datasets, we generated a benign dataset of \num{1430} packages.
This methodology follows the practices of Zahan et al.~\cite{zahan2022weak}.  

It is possible that popular packages are different from a typical Python package: potentially better engineered and more conformant to standard Python programming practices.
Consequently, using only popular packages as the benign dataset might lead to unrealistic benchmark results since these packages might be relatively easy for detection tools to classify as benign~\cite{wright2020hunting} .

\textit{Random PyPI packages:} The second step was to select the most recent version of \num{1000} randomly chosen Python packages.
We excluded the packages that lacked any Python files and any packages that were no longer available at the time of analysis.
This led to a benign dataset of \num{986} packages.
While there is a chance that some of these packages are malicious, the chance that more than a handful of these packages are malicious is small.
Conservatively assuming \num{2000} undetected malicious packages on PyPI (of \num{400000} total), the probability that more than \num{10} packages in this set are malicious is less than 2 percent; this bounds the potential error in our false positive and true negative rates at around \SI{1}{\percent}.

While we were unable to audit the \num{45} million lines of code in the popular packages dataset in \Cref{tab:descriptive-statistics}, we did perform spot checks on a small sample of benign packages flagged as malware.
For instance, the ostensibly benign popular packages \texttt{pymupdf} and \texttt{configargparse} were flagged as malicious by one of the scanners. None of the spot checks revealed malicious code, although they contained suspicious patterns.

\Cref{tab:descriptive-statistics} details the number of packages in each benchmark sub-dataset.
Note that for each package, we obtain the latest release. The number of Python files and the number of lines of code of the popular packages are significantly larger than the random and malicious packages because popular packages tend to provide much more functionality.

\begin{table}
  \centering
  \caption{Descriptive Statistics of the Benchmark Dataset}
  \longcaption{\columnwidth}{Line counts from \texttt{scc}~\cite{scc-tool}. We use only the latest release of each package.}
  \label{tab:descriptive-statistics}
  \begin{tabular}{@{}lS[table-format=4]S[table-format=6]S[table-format=8]@{}}
    \toprule
    \textbf{Dataset}  &  \textbf{Packages} & \textbf{Python Files} & \textbf{Lines of Code} \\
    \midrule
    Malicious        & 168 & 1339 & 228192  \\
    Benign (popular)   & 1430 & 164223 & 45254876 \\
    Benign (random)   & 986 & 16832 & 2770978\\
    \bottomrule
  \end{tabular}
\end{table}

\begin{table*}[ht]
\caption{Packages with at least one alert, by tool (\%).}
\label{tab:alerts}
\centering
\begin{tabular}{@{}lS[table-format=2.1]S[table-format=2.1]S[table-format=2.1]S[table-format=2.1]S[table-format=2.1]S[table-format=2.1]@{}}
\toprule
 & \multicolumn{2}{c}{PyPI checks}              & \multicolumn{2}{c}{OSS Detect Backdoor}              & \multicolumn{2}{c}{Bandit4Mal}                       \\ \cmidrule(lr){2-3}  \cmidrule(lr){4-5} \cmidrule(l){6-7}
\textbf{Dataset}                  & {\texttt{setup.py} (\%)} & {\texttt{*.py} (\%)} & {\texttt{setup.py} (\%)} & {\texttt{*.py} (\%)} & {\texttt{setup.py} (\%)} & {\texttt{*.py} (\%)} \\ \midrule
Malicious                & 58.9        & 85.7           & 50.6        & 85.1           & 66.7        & 90.5           \\
Benign (popular)           & 33.2        & 94.2           & 41.7        & 96.6           & 79.5        & 96.6           \\
Benign (random)            & 14.9        & 67.8           & 33.1        & 77.2           & 70.2        & 84.6           \\ \bottomrule
\end{tabular}
\end{table*}

\subsection{Benchmarking Evaluation}\label{subsec:benchmarking-evaluation}

\noindent
\Cref{tab:alerts} reports the number of packages with one or more alerts for all datasets using all approaches described in \Cref{sec:evaluation}. In this section, we report the results of the analysis of the alerts generated by the malware detection tools.

\paragraph{Scanners catch more than half of malicious packages}
All checks, when run on only \texttt{setup.py} files, fired alerts (true positive) for over \SI{50}{\percent} of packages, with OSS detect backdoor the lowest (\SI{50.6}{\percent}) and Bandit4Mal the highest (\SI{66.7}{\percent}).
When including all Python files, the checks detected over \SI{85}{\percent} of malicious packages.
All approaches, no matter the test corpus, had true positive rates above \SI{50}{\percent}.

\paragraph{False positive rates are high (sometimes higher than true positive rates)}
Using a single alert as our threshold, the lowest false positive rates were found checking only \texttt{setup.py} files.
However, we found false positive rates of at least \SI{14.9}{\percent} (PyPI malware checks, random packages) and as high as \SI{79.5}{\percent} (Bandit4Mal, popular packages).

The false positive rate increases when checking all files (since the alerts that fire are a superset of those for the \texttt{setup.py} alerts).
The positivity rate was highest for the popular packages, which tended to have more files and more lines of code than the malicious or random packages.
Bandit4Mal and OSS Detect Backdoor had the greatest false positive rates (\SI{96.6}{\percent}) when scanning all Python files.
In fact, the false positive rate for popular packages with these tools was \emph{higher} than the true positive rate for malicious packages.

\paragraph{Positive packages had many alerts} 
These checks may fire multiple alerts per package, and in many cases they did.
A user of these checks might have to investigate all such alerts to render a verdict on a suspicious package.

Scanning only \texttt{setup.py} files, we find that all methods have a median of \num{3} or fewer alerts among the benign packages which triggered \emph{any} alerts.
However, the maximum number of alerts among benign packages (which triggered at least one alert) was high: \num{45}, \num{77}, and \num{75} for the PyPI checks, OSS Detect Backdoor, and Bandit4Mal, respectively.
For malicious packages, the numbers were not necessarily higher: the maximum number of alerts was \num{3} for the PyPI malware checks.

When scanning all Python files, the number of alerts increases dramatically.
The median number of alerts (for popular benign packages packages with at least one alert) was \num{10}, \num{85}, and \num{19} for the PyPI checks, OSS Detect Backdoor, and Bandit4Mal, respectively.
The PyPI checks fired \num{5377} alerts for one popular package, and Bandit4Mal fired \num{16155} alerts for one as well.
But the noisiest was OSS Detect Backdoor, which fired \num{124267} for one benign (popular) package.

\paragraph{Increasing alerting thresholds drops true positive rates while false positive rates remain nontrivial}
The above results may have over-estimated false positive rates, since we used the lowest possible threshold of one alert.
It is possible that some other threshold could lead to superior performance results of the tools in terms of balancing the tradeoff between the number of true positives and false positives.

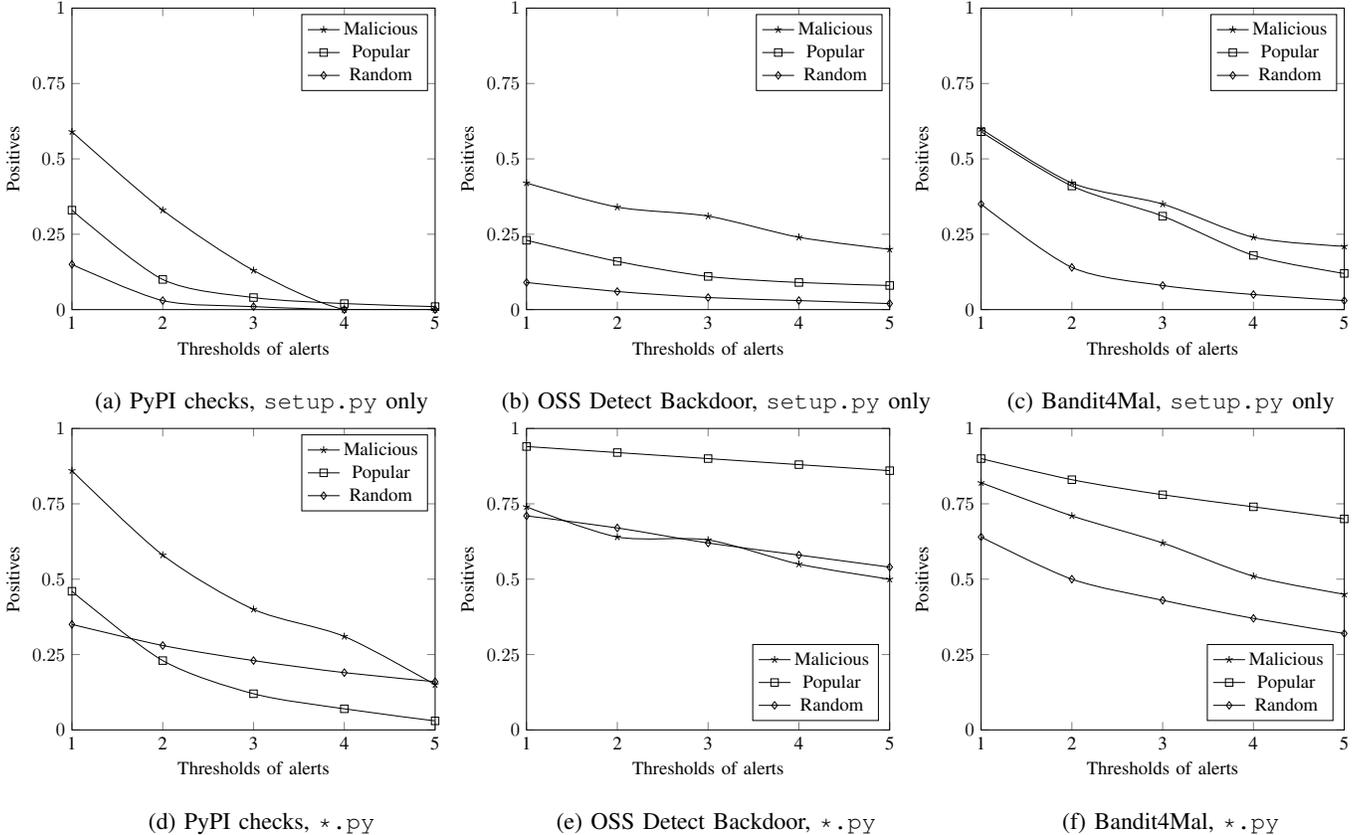
\begin{figure*}[ht]
\begin{subfigure}[t]{0.15\textwidth}
\resizebox{2.4in}{!}{
\begin{tikzpicture}
\begin{axis}[
    xlabel=Thresholds of alerts,
    ylabel=Positives,
    xmin=1, xmax=5,
    ymin=0, ymax=1,
    xtick={1,2,3,4,5},
    xticklabels={1,2,3,4,5},   
    ytick={0,0.25,0.5,0.75,1}
            ]
\addplot[smooth,mark=star] plot coordinates {
    (1,0.59)
    (2,0.33)
    (3,0.13)
    (4, 0)
    (5, 0)
};
\addlegendentry{Malicious}

\addplot[smooth,mark=square] plot coordinates {
    (1,0.33)
    (2,0.1)
    (3,0.04)
    (4, 0.02)
    (5, 0.01)
};
\addlegendentry{Popular}

\addplot[smooth,mark=diamond] plot coordinates {
    (1,0.15)
    (2,0.03)
    (3,0.01)
    (4, 0)
    (5, 0)
};
\addlegendentry{Random}
\end{axis}
\end{tikzpicture}
}
\captionsetup{justification=centering,margin=-2.2cm}
\caption{PyPI checks, \texttt{setup.py} only} \label{fig:setup-pypi}
\end{subfigure}\hspace{32mm}
\begin{subfigure}[t]{0.15\textwidth}
\resizebox{2.4in}{!}{
\begin{tikzpicture}
\begin{axis}[
    xlabel=Thresholds of alerts,
    ylabel=Positives,
    xmin=1, xmax=5,
    ymin=0, ymax=1,
    xtick={1,2,3,4,5},
    xticklabels={1,2,3,4,5},   
    ytick={0,0.25,0.5,0.75,1}
            ]
\addplot[smooth,mark=star] plot coordinates {
    (1,0.42)
    (2,0.34)
    (3,0.31)
    (4, 0.24)
    (5, 0.2)
};
\addlegendentry{Malicious}

\addplot[smooth,mark=square] plot coordinates {
    (1,0.23)
    (2,0.16)
    (3,0.11)
    (4, 0.09)
    (5, 0.08)
};
\addlegendentry{Popular}

\addplot[smooth,mark=diamond] plot coordinates {
    (1,0.09)
    (2,0.06)
    (3,0.04)
    (4, 0.03)
    (5, 0.02)
};
\addlegendentry{Random}
\end{axis}
\end{tikzpicture}
}
\captionsetup{justification=centering,margin=-2.2cm}
\caption{OSS Detect Backdoor, \texttt{setup.py} only} \label{fig:setup-odb}
\end{subfigure}\hspace{32mm}
\begin{subfigure}[t]{0.15\textwidth}
\resizebox{2.4in}{!}{
\begin{tikzpicture}
\begin{axis}[
    xlabel=Thresholds of alerts,
    ylabel=Positives,
    xmin=1, xmax=5,
    ymin=0, ymax=1,
    xtick={1,2,3,4,5},
    xticklabels={1,2,3,4,5},   
    ytick={0,0.25,0.5,0.75,1}
            ]
\addplot[smooth,mark=star] plot coordinates {
    (1,0.6)
    (2,0.42)
    (3,0.35)
    (4, 0.24)
    (5, 0.21)
};
\addlegendentry{Malicious}

\addplot[smooth,mark=square] plot coordinates {
    (1,0.59)
    (2,0.41)
    (3,0.31)
    (4, 0.18)
    (5, 0.12)
};
\addlegendentry{Popular}

\addplot[smooth,mark=diamond] plot coordinates {
    (1,0.35)
    (2,0.14)
    (3,0.08)
    (4, 0.05)
    (5, 0.03)
};
\addlegendentry{Random}
\end{axis}
\end{tikzpicture}
}
\captionsetup{justification=centering,margin=-2.2cm}
\caption{Bandit4Mal, \texttt{setup.py} only} \label{fig:setup-b4m}
\end{subfigure}

\begin{subfigure}[t]{0.15\textwidth}
\resizebox{2.4in}{!}{
\begin{tikzpicture}
\begin{axis}[
    xlabel=Thresholds of alerts,
    ylabel=Positives,
    xmin=1, xmax=5,
    ymin=0, ymax=1,
    xtick={1,2,3,4,5},
    xticklabels={1,2,3,4,5},   
    ytick={0,0.25,0.5,0.75,1}
            ]
\addplot[smooth,mark=star] plot coordinates {
    (1,0.86)
    (2,0.58)
    (3,0.4)
    (4, 0.31)
    (5, 0.15)
};
\addlegendentry{Malicious}

\addplot[smooth,mark=square] plot coordinates {
    (1,0.46)
    (2,0.23)
    (3,0.12)
    (4, 0.07)
    (5, 0.03)
};
\addlegendentry{Popular}

\addplot[smooth,mark=diamond] plot coordinates {
    (1,0.35)
    (2,0.28)
    (3,0.23)
    (4, 0.19)
    (5, 0.16)
};
\addlegendentry{Random}
\end{axis}
\end{tikzpicture}
}
\captionsetup{justification=centering,margin=-2.2cm}
\caption{PyPI checks, \texttt{*.py}} \label{fig:all-pypi}
\end{subfigure}\hspace{32mm}
\begin{subfigure}[t]{0.15\textwidth}
\resizebox{2.4in}{!}{
\begin{tikzpicture}
\begin{axis}[
    xlabel=Thresholds of alerts,
    ylabel=Positives,
    legend pos=south east,
    xmin=1, xmax=5,
    ymin=0, ymax=1,
    xtick={1,2,3,4,5},
    xticklabels={1,2,3,4,5},   
    ytick={0,0.25,0.5,0.75,1}
            ]
\addplot[smooth,mark=star] plot coordinates {
    (1,0.74)
    (2,0.64)
    (3,0.63)
    (4, 0.55)
    (5, 0.5)
};
\addlegendentry{Malicious}

\addplot[smooth,mark=square] plot coordinates {
    (1,0.94)
    (2,0.92)
    (3,0.9)
    (4, 0.88)
    (5, 0.86)
};
\addlegendentry{Popular}

\addplot[smooth,mark=diamond] plot coordinates {
    (1,0.71)
    (2,0.67)
    (3,0.62)
    (4, 0.58)
    (5, 0.54)
};
\addlegendentry{Random}
\end{axis}
\end{tikzpicture}
}
\captionsetup{justification=centering,margin=-2.2cm}
\caption{OSS Detect Backdoor, \texttt{*.py}} \label{fig:all-odb}
\end{subfigure}\hspace{32mm}
\begin{subfigure}[t]{0.15\textwidth}
\resizebox{2.4in}{!}{
\begin{tikzpicture}
\begin{axis}[
    xlabel=Thresholds of alerts,
    legend pos=south east,
    ylabel=Positives,
    xmin=1, xmax=5,
    ymin=0, ymax=1,
    xtick={1,2,3,4,5},
    xticklabels={1,2,3,4,5},   
    ytick={0,0.25,0.5,0.75,1}
            ]
\addplot[smooth,mark=star] plot coordinates {
    (1,0.82)
    (2,0.71)
    (3,0.62)
    (4, 0.51)
    (5, 0.45)
};
\addlegendentry{Malicious}

\addplot[smooth,mark=square] plot coordinates {
    (1,0.9)
    (2,0.83)
    (3,0.78)
    (4, 0.74)
    (5, 0.7)
};
\addlegendentry{Popular}

\addplot[smooth,mark=diamond] plot coordinates {
    (1,0.64)
    (2,0.5)
    (3,0.43)
    (4, 0.37)
    (5, 0.32)
};
\addlegendentry{Random}
\end{axis}
\end{tikzpicture}
}
\captionsetup{justification=centering,margin=-2.2cm}
\caption{Bandit4Mal, \texttt{*.py}} \label{fig:all-b4m}
\end{subfigure}\hspace{32mm}
\caption{Percentage of packages with at least the given threshold of alerts.}
\label{fig:positives-alerts}
\end{figure*}

\Cref{fig:positives-alerts} uses a variety of alert thresholds to plot the true positive rate for the malicious packages and the false positive rates for both benign package datasets.
This figure contains two sub-graphs per detection tool, one for the rates when scanning only \texttt{setup.py} files and another when all Python files are scanned.
In general, \Cref{fig:positives-alerts} suggests that a low threshold of alerts generated many false positives, especially when using Bandit4Mal and OSS Detect Backdoor on all Python files.
On the other hand, setting a threshold higher than 4 alerts makes the tools report fewer malicious packages, or even none in the case of PyPI malware checks on \texttt{setup.py} files.
Across all the tools, we observed that \num{3} alerts seems to be an optimal threshold for the tools to balance the number of malicious packages identified and false positives.  


\begin{figure}[t]
    \centering
    \resizebox{3.6in}{!}{
\begin{tikzpicture}
  \begin{axis}[
    xbar,
    bar width=0.2cm,
    y axis line style = { opacity = 0 },
    axis x line       = none,
    tickwidth         = 0pt,
    enlarge y limits  = 0.2,
    enlarge x limits  = 0.02,
    symbolic y coords = {Deserialization, Metaprogramming, Networking, Process Spawn, Subprocess},
    nodes near coords,
  ]
  \addplot[pattern=horizontal lines] coordinates { (10.3,Deserialization)  (27.4,Metaprogramming)
                (35.4,Networking) (8.6,Process Spawn) (18.3,Subprocess)};
  \addplot[pattern=north east lines] coordinates { (0.3,Deserialization)  (67.4,Metaprogramming)
                (1.4,Networking) (16.8,Process Spawn) (14.1,Subprocess)};
  \addplot[pattern=dots] coordinates { (0.5,Deserialization)  (68,Metaprogramming)
                (3,Networking) (18.8,Process Spawn) (9.6,Subprocess)};
  \legend{Malicious, Popular, Random}
  \end{axis}
\end{tikzpicture}
}
    \caption{Distribution of the rules in the three datasets}
    \vspace{0.25cm}
    \longcaption{\columnwidth}{Note: This analysis analyzed only \texttt{setup.py} files for each package. The x-axis is on log-scale.
}
    \label{fig:rules-distributions}
\end{figure}
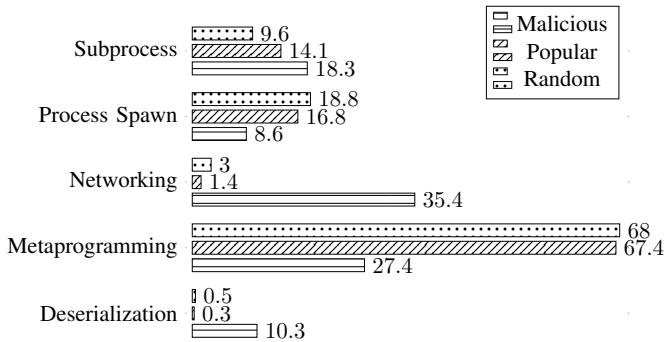

\paragraph{Some rules are more effective than others}
To understand why these tools flagged so many benign packages as malicious, we broke down the specific rules that were triggered in the case of the PyPI malware checks.
\Cref{fig:rules-distributions} shows the distribution of the alerts for each rule in the \texttt{setup.py} files of the three datasets.
We observed that \texttt{metaprogramming\_in\_setup} is the most common rule triggered in the popular and random packages.
However, malicious packages contain the highest percentage of \texttt{networking\_in\_setup} alerts.
This indicates the indicators of a networking event could provide a higher confidence of maliciousness.
We found a similar frequency of the \texttt{process\_spawn\_in\_setup} and \texttt{subprocess\_in\_setup} rules in all the datasets, which suggests that using this rule is not so effective in distinguishing malicious and benign code.  

\paragraph{The tools ran in a reasonable amount of time, especially for \texttt{setup.py} files}
Tested on a laptop with \num{4} CPU cores and \num{8} GB RAM,
the PyPI checks and OSS Detect Backdoor ran with a median time around \num{2} seconds per package for all datasets and a maximum run time of \num{26} minutes for one popular package (the package \texttt{ansible}). Also, we observed that OSS Detect Backdoor ran slightly slower than PyPI malware checks (\num{0.8} seconds, on average). Bandit4Mal took less than a second to process a malicious or random package, on average. However, the tool runs with a median time of nearly \num{5} seconds for the popular package dataset and a maximum run time of \num{2.5} hours for one popular package.  

\section{Discussion} \label{sec:discussion}
\noindent
In our qualitative research (\cref{sec:interviews}), we find that the motivation and even the setting of malware detection for package repositories differ in practice from those assumed in the literature.
In particular, removing all malware from PyPI is a low priority for maintainers relative to other past, planned, and ongoing security interventions, as many malicious packages are downloaded very rarely, and consequently have minimal impact on real developers.
Based on this, we formulate requirements for an automated malware scanner deployed in this setting.
In our quantitative research (\cref{sec:evaluation}), we investigate existing tools in the context of these requirements, evaluating both the performance (with a particular focus on false positive rates across tunable thresholds) and ease of deployment.
We find that the tools we evaluate do not meet these requirements.
However, these tools may still be useful in the broader repository malware detection ecosystem.

\subsection{Requirements for repository-side automated malware scanning}
\noindent
There are two types of repository-side malware checks: \emph{blocking}, in which a published package is not accepted until after a scan, or \emph{passive}, in which releases are scanned after-the-fact.
Both require ongoing intervention by repository maintainers, who are often volunteers or employees of a nonprofit and under-resourced.
For any blocking checks, there must be an appeal process.
For passive checks, positive results trigger alerts to maintainers; after reviewing alerts, the maintainers remove any packages they deem malicious.
The following requirements apply to both settings.

\textbf{``Zero'' false-positives.}
Malicious packages are relatively rare; consequently, even with very low false positive rates (\SI{0.1}{\percent}) and a perfect true positive rate (an unrealistic goal with current methods), the majority of alerts for an automated system will be false positives.
Maintainers are unwilling to spend more than a minute or so per alert (ideally even less), or review alerts in real-time (for instance, in the middle of the night).
The ideal tool would have a ``zero'' false-positive rate: at most \SI{0.001}{\percent} for a passive check, and even lower for a blocking one.

\textbf{Low effort to adopt and maintain.}
Engineering and maintenance effort limit the checks that can be deployed: static checks, which do not require running potentially malicious code, are feasible, while dynamic checks, which must run in a carefully-built sandbox, are not.
Similarly, interviewees expressed skepticism about ``research-grade'' software, which they view as insufficiently robust for a load-bearing role in their repository.

\textbf{Low-effort to use.}
Administrators want to make very few value judgments, and prefer their interaction with a report of malware to involve only a quick confirmation that the code looks malicious.
Checks should have a binary decision (possibly based on a tunable threshold) and present a guess about whether a package is malicious, not just a list of suspicious features.
However, checks should expose evidence of malicious behavior, as this allows administrators to confirm before taking the package down. Fortunately, this setting allows trade-offs.

\textbf{False negatives are acceptable.}
Repository administrators believe that no automated system can detect all or even most attacks. However, such systems are not useless: PyPI sees a steady, large volume of low-effort attacks, and even a system with limited sensitivity would help to clean these up.

\textbf{Computing resources are generous.}
Any blocking checks must run in at most seconds, since users expect packages to be available immediately after upload.
However, passive checks can take much longer.
Further, PyPI has access to a large amount of cloud computing resources, and could deploy even computationally expensive passive checks.

\subsection{Current techniques fall short}
\noindent
The tools we evaluated (\cref{sec:evaluation}) had several traits which immediately disqualify them for inclusion in a repository malware analysis pipeline.
They emitted alerts for between \SI{15}{\percent} and \SI{97}{\percent} of benign packages.
These tools are flexible: they support user-supplied suspicious patterns and allow users to make their own decision rules.
In this context, however, flexibility is a fault: it increases the burden of maintenance and decision making for repository administrators. 

These tools did have some advantages.
They were all static checks and reasonably easy to run in an automated pipeline.
They ran in times appropriate for passive checks, though not blocking checks.
Overall, such tools are unsuitable for deployment as repository-side malware scanners.
However, they may still be useful elsewhere.

\subsection{Alternative: a socio-technical malware detection system}
\noindent
There \emph{are} malware checks for PyPI, but they involve a symbiotic ecosystem comprising people and organizations, not just computers.
For-profit companies and research institutions gain reputation and prestige by finding malware; they report this to the maintainers in exchange for the opportunity to publicize their work.
Maintainers outsource malware detection to researchers who are willing to troll through numerous false alerts in order to evaluate their methods or claim credit for discovery.
This system works well enough to allow maintainers to focus on other security work.

In such a system, the onus falls on researchers to find malware.
These researchers develop and evaluate new tools.
They also manually deal with false positives.
Such a researcher may find a \SI{15}{\percent} false positive rate acceptable, at least in the short term, and they can use any number of techniques in their toolkit for malware hunting.

In some ways, this ``solution'' mirrors broader trends in open-source software related to the involvement of companies in open-source software and the ability of disparate parties to coordinate and divide tasks via online communication~\cite{geer2020will, schweik2012internet}.
Nonetheless, there is likely still ample room, according to all interviewees, to reduce transaction costs and improve collaboration between PyPI, other OSS registries, enterprises, and individuals scanning these repositories for malware.

\section{Recommendations} \label{sec:recommendations}
\noindent
We conclude this paper with recommendations for academics and security researchers to improve the quality and usability of OSS malware detection tools, especially for PyPI.

\textbf{Most effort should be elsewhere.}
Our interviewees reported long lists of proposed security improvements and limited ability to implement these.
Some are a poor fit for researchers: they require money, time, or large-scale (but not particularly novel) engineering projects.
But others require skills that security researchers have: threat modeling, systems design, and analysis, or cryptographic implementation.

\textbf{Researchers should design for malware detection as practiced.}
The requirements laid out above rule out even the very best academic tools.
Efforts to meet these requirements are likely futile.
However, giving up on this ambition frees researchers to experiment with the tools that work for them: they no longer need to write purely non-interactive scanners or even aspire to low false-positive rates.
Designing tools for external researchers removes constraints and allows experimental methods.

\textbf{Improve the socio-technical system.}
The ecosystem which has organically emerged to deal with the threat of malware on PyPI involves a number of organizations and individuals, each with their own goals.
Perhaps by accident, the goals of these individuals are mostly aligned; a system deliberately designed, for instance, in collaboration with researchers could have even better outcomes.
However, administrators were wary of perverse incentives: monetary rewards for reporting malware may inspire unscrupulous researchers to plant malware for themselves to ``find.''

\textbf{Consider the goals of repository administrators.}
Removing all malware is a low priority for administrators.
They worry much more about critical packages~\cite{sharma2022pypi} or packages that users may accidentally install via a typo.
While suspicious code by itself is too weak of a signal on which to remove a package, suspicious code in a suspiciously-named package might be.
Combined metadata and behavior-based indicators may be a more useful signal than either on their own.
Similarly, administrators prioritize low false positive rates over catching all or most attacks; this may indicate that signature-based malware detection is, in fact, appropriate, despite the trend in the academic literature over the past decades to the contrary.

\textbf{Focus on usability.}
Existing systems emit long lists of ``alerts,'' which a human user must investigate and interpret.
Instead, the output should be actionable: a concrete prediction that a package warrants further investigation.
Additionally, further investigation should be as easy and useful as possible.
Research on human-computer interaction may help create tools that give useful signals while still allowing researchers or administrators to quickly drill down and root out malware.

\textbf{Open science principles facilitate science.}
We omitted several tools because their detection rules were missing, or their source was unavailable.
Researchers should strive for reproducibility, shared datasets, and consistent interfaces which make cross-tool comparison easy.
We make a minor contribution in this area by publishing metadata that allows easily reconstructing our dataset.

\textbf{Funders should embrace this ecosystem.}
While administrators maintain relationships with researchers and act on their guidance, they have a number of shovel-ready ideas~\cite{psf2022productionize} to make this system more efficient.
For instance, PyPI does retain all packages removed as malware, but the metadata is unavailable.
Making such a dataset available to identified researchers would enable training and honing much better research tools.
Similarly, repositories should facilitate scanning via a ``firehose''-like stream of updates or bulk datasets of all packages (for instance, torrents).
Companies who have built businesses on open-source software, nonprofits, and governments should contribute engineering resources or funding to enable implementation.

\section{Threats to Validity}
\textbf{The malicious dataset may not accurately represent malicious packages in the wild.}
Not all malicious PyPI packages are publicly known.
To the best of our knowledge, Backstabber’s Knife Collection and MalOSS samples are the largest OSS supply-chain malware repositories available for researchers upon request.
However, our interviewees report removing about \num{100} malicious packages per month, suggesting that these datasets capture only a small portion of total PyPI malware.
The \num{168} packages in the analyzed dataset likely exhibit some sampling bias, since they were those that researchers were able to track down; a randomly-sampled dataset of malware might contain samples with different characteristics.

\textbf{The benchmarking analysis excluded a number of OSS malware detection tools.}
Several other OSS malware detection tools exist but were excluded from the study either because the tool had unavailable source code, did not use behavior-based analysis, or lacked published detection rules.
Also, there are OSS malware detection tools designed for other programming language ecosystems (e.g., \cite{ohm2022feasibility} for npm) that could potentially be used, after engineering modifications, to analyze PyPI packages.
Future evaluations could therefore benchmark more and different OSS malware detection tools.

\textbf{More sophisticated decision rules might improve performance.}
This analysis used decision rules related to only the number of alerts for each package, finding that false positive rates were unacceptably high even when using a threshold that eliminated most true positives.
More sophisticated rules might use signals like the type of alert to deliver better performance.

\textbf{This evaluation only analyzed the latest versions of the packages.}
Most attackers target the latest versions of a package, which maximizes the number of victims, so the benchmark analysis only examined the latest version of each package.
Some attackers, however, inject malicious code into older versions of a package, possibly hoping that the developers who use older versions are less cautious about security~\cite{duan2021towards}.


\section{Conclusion and Future Work}

This paper's benchmarking of approaches to Python malware detection in a repository setting resulted in a clear finding: high false positive rates uniformly characterize these approaches, which imposes a high burden on any single organization that decides to police PyPI from malware. 

Fortunately, the interview component of this paper revealed that a socio-technical system involving both repository maintainers and security researchers jointly engaged in PyPI malware detection has emerged. This coalition enables security researchers, who have a higher tolerance for false positives than repository administrators, to specialize in the malware detection component of the system. This arrangement is a feature, not a bug. Parties interested in strengthening repository defenses against malware, at PyPI and potentially other OSS repositories too, should seek to strengthen this system and reduce the coordination and transaction costs between the different parties.

\bibliographystyle{IEEEtran}
\bibliography{main}
\end{document}